\documentclass[aps,showpacs,manuscript,12pt]{revtex4}
\usepackage{amssymb}
\usepackage{amsmath}
\usepackage{graphicx}

\begin{document}
\title{\textbf{Inverse kinetic theory approach to turbulence theory$^{\S }$}}
\author{M. Tessarotto$^{a,b}$, M. Ellero$^{c}$ and P.
Nicolini $^{a,b}$} \affiliation{\ $^{a}$Department of Mathematics
and Informatics, University of Trieste, Italy, $^{b}$Consortium of
Magneto-fluid-dynamics, University of Trieste, Italy,
$^{c}$Department of Aerodynamics, Technical University of Munich,
Munich, Germany}
\begin{abstract}
A fundamental aspect of turbulence theory is related to the
identification of realizable phase-space statistical descriptions
able to reproduce in some suitable sense the stochastic fluid
equations of a turbulent fluid. In particular, a major open issue
is whether a purely Markovian statistical description of
hydrodynamic turbulence actually can be achieved. Based on the
formulation of a \textit{deterministic inverse kinetic theory}
(IKT) for the 3D incompressible Navier-Stokes equations, here we
claim that such a \textit{Markovian statistical description
actually exists}. The approach, which involves the introduction of
the \textit{local velocity probability density} for the fluid
(local pdf) - rather than the velocity-difference pdf adopted in
customary approaches to homogeneous turbulence - \ relies
exclusively on first principles. These include - in particular -
the exact validity of the stochastic Navier-Stokes equations, the
principle of entropy maximization and a constant H-theorem for the
Shannon statistical entropy. As a result, the new approach affords
an exact equivalence between Lagrangian and Eulerian formulations
which permit local pdf's which are generally non-Maxwellian (i.e.,
non-Gaussian). The theory developed is quite general and applies
in principle even to turbulence regimes which are non-stationary
and non-uniform in a statistical sense.
\end{abstract}
\maketitle



\section{Introduction}

In this and in an accompanying paper \cite{Tessarotto2008-aa} the
possibility of formulating inverse kinetic theory (IKT) approaches
to hydrodynamic turbulence theory is investigated. By definition,
these are meant to be as phase-space models able to deliver a
prescribed set \ (or subset) of fluid equations to be expressed in
terms of appropriate moment equations of a suitable statistical
equation, denoted as inverse kinetic equation (IKE). Depending of
the subset of fluid equations to be considered different (and
possibly non-unique) IKT approaches can\ in principle be
developed. In the present paper, in particular, a Markovian
statistical
model of turbulence is obtained, based on the formulation of a \emph{%
deterministic inverse kinetic equation} for the
stochastic-averaged Navier-Stokes equations. The theory permits
the explicit construction of the \emph{local position-velocity
joint probability distribution function} (local pdf) which
advances in time the corresponding stochastic-averaged \ fluid
fields and which can be shown to characterize uniquely an
incompressible isothermal fluid.

\subsection{The unsolved problem}

The theoretical approach to the turbulence problem in
incompressible fluids is one of the outstanding intellectual
challenges of \ contemporary physics. Turbulence theory has been
pioneered by Kolmogorov (K41,\cite{Kolm}) who - using simple
heuristic arguments based on dimensional analysis - first shed
light on the understanding homogeneous turbulence (HT, i.e.,
stationary, spatially homogeneous and isotropic turbulence). This
led to the interpretation of HT as a self-similar energy cascade
in which turbulent eddies transport the kinetic energy of the
fluid from a prescribed injection scale to a suitable dissipation
scale (the so-called inertial range). This lead to the well-known
Kolmogorov "5/3-power law" conjecture \cite{Kolm} - later
confirmed by experiments and numerical results \cite{experiment} -
that in the inertial range the cascade is characterized by energy spectrum $%
E(k)$ of the form $E(k)=K_{Ko}\Pi ^{2/3}k^{-5/3},$ where $K_{Ko}$
is an
universal constant called Kolmogorov's constant, $k$ is the wavenumber, and $%
\Pi $ is the nonlinear cascade of energy, to be identified with
the dissipation rate of the fluid. \ Subsequently the problem of
homogeneous turbulence was approaches with the goal of quantifying
the processes underlying the spectrum. Various theories were
developed, for this purpose, to understand fluid turbulence, based
on attempts to introduce suitable models for the statistics of
turbulent flows \cite{Monin1975,Pope2000}. These phenomenological
theories can be developed, in principle, choosing for the
description of fluids either Eulerian or Lagrangian viewpoints.
The two approaches, if fluid dynamics were fully understood,
should be completely equivalent. Unfortunately, at least for the
treatment of turbulent fluids, we are still quite far from
reaching this goal. The main historical reason of this situation
can be understood by looking at the customary statistical approach
based on so-called velocity probability density function (pdf)-
method for an incompressible fluid (for a review see for example \cite%
{Pope2000}). In the Lagrangian treatment of turbulence [denoting
the
so-called \emph{Lagrangian turbulence }(LT)] the Lagrangian path $\mathbf{R}(%
\mathbf{r,}t)$ and the velocity $\mathbf{U}(\mathbf{r,}t)$ of a
fluid element, initially starting at the position $\mathbf{r,}$
are determined by the equations:
\begin{eqnarray}
\frac{d\mathbf{R}(\mathbf{r,}t)}{dt} &=&\mathbf{U}(\mathbf{r,}t),
\label{velocity} \\
\frac{d\mathbf{U}(\mathbf{r,}t)}{dt}
&=&\mathbf{A}(\mathbf{r,}t;g), \label{accelle}
\end{eqnarray}%
where the vector field $\mathbf{A}(\mathbf{r,}t;g),$ to be
suitably related to the Navier-Stokes equations, \ is assumed to
depend on an appropriate statistical distribution $g.$ Usually
this is identified with the joint position-velocity probability
distribution (pdf) of the particle for the
increments $\mathbf{x}(t)=\mathbf{R}(t,\mathbf{y})-\mathbf{U}(0,t)t-\mathbf{%
r,}$
$\mathbf{u}(t)=\mathbf{U}(t,\mathbf{y})-\mathbf{U}(0,\mathbf{y})$
defined as
\begin{equation}
\left\{
\begin{array}{c}
g(\mathbf{x,u},t)=\left\langle \delta (\mathbf{x}-\mathbf{x}(t))\delta (%
\mathbf{u}-\mathbf{u}(t))\right\rangle \\
f(\mathbf{x,u},0)=\delta (\mathbf{x})\delta (\mathbf{u}).%
\end{array}%
\right. ,  \label{pdf-1}
\end{equation}%
where the brackets denote a suitable stochastic average over a
stationary statistical ensemble. For HT, the velocity increment
pdf (\ref{pdf-1}) can be shown to obey a Fokker-Planck statistical
equation \cite{Pope2000}. The fundamental reason why this happens
is that a statistical theory of turbulence, relying solely of
first principles, is yet not available. For this reason in the
past various statistical models, based on heuristic assumptions
about the statistics of $\mathbf{A}(t,\mathbf{y;}g)$ and of the
related pdf $g$, have been introduced. In turbulence theory $g$ is
usually identified with the velocity-difference probability
density function (pdf), traditionally adopted for the description
of homogeneous turbulence. In these cases, however, the
corresponding Fokker-Planck equation for $g$ generally does not
define a closed system of moment equations (which should actually
coincide with the Navier-Stokes equations themselves). This raises
the critical issue of the closure of the statistical models of
fluid turbulence and the difficulty of achieving equivalent
Eulerian and Lagrangian formulations. In the past several
statistical models have been proposed for the determination of
$g$, which include the mapping-closure
method, the test-function method and the field-theoretical approach \cite%
{Chen1989}. However, a critical aspect to turbulence theory is the
possible appearance of coherent structures (like vortices and
convective cells). In fluid turbulence the signature of the
presence of coherent structures is provided by the existence of
non-Gaussian features in the probability density. During the last
few years lots of efforts have been put into the formulation of
more sophisticated phenomenological theories which can take into
account these facts (c.f. the review article \cite{Aringazin}).
These approaches are based on a theoretical analysis of the
Navier-Stokes equation or on the advanced data analysis of the
experimentally obtained Lagrangian path's of particles (see for
example \cite{Friedrich}). Nevertheless, despite the progress
achieved in modelling key features of the basic phenomenology,
still missing is a consistent, theory-based, statistical
description of fluid turbulence. Clearly, such formulation - if
achievable at all - should rely exclusively on a rigorous,
deductive formulation of the
turbulence-modified fluid equations following from the fluid equations \cite%
{Sreenivasan1997}. Based on a recently proposed inverse kinetic
theory (IKT)
for incompressible fluids (Ellero et al., 2004-2008 \cite%
{Ellero2000,Ellero2005,Tessarotto2006,Tessarotto2008}), here we
intend to formulate a deterministic IKT for the
stochastic-averaged Navier-Stokes equations. \ Key features of the
new theory are: \emph{1) the formulation is based on the
\underline{\emph{local pdf}}, rather than the velocity-increment
pdf usually adopted in turbulence theory (see for example
\cite{Pope2000}); 2) the pdf is advanced in time by means of a
Markovian Vlasov-type kinetic equation; \ 3) the kinetic equation
implies the exact validity of the stochastic-averaged
Navier-Stokes equations (see below) and as a consequence the
kinetic equation satisfies exact closure conditions;} \emph{4) the
theory displays a complete equivalence of the Lagrangian and
Eulerian viewpoints \cite{Tessarotto2008}; 5) the theory is based
on first principles, i.e., besides the Navier-Stokes equations,
the validity of the principle of entropy maximization and a
constant H-theorem to be imposed on the Shannon statistical
entropy.} The theoretical setting here adopted is based on the
definition of a "restricted" phase-space representation of the
fluid \cite{Tessarotto2008}, whereby the phase space ($\Gamma )$
is
identified with the direct-product space $\Gamma =\Omega \times V,$ where $%
\Omega ,V\subseteq $ $%
\mathbb{R}
^{3},$ $\Omega $ is an open set which coincides with the fluid
domain (i.e.,
the bounded sub-domain of $%
\mathbb{R}
^{3}$ occupied by the fluid) and $V$ is the velocity space.

\section{Stochastic-averaged fluid equations}

For definiteness here we consider an incompressible isothermal
fluid immersed in a fluid domain $\Omega $ (to be identified with
a bounded
subdomain of $%
\mathbb{R}
^{3}$ with closure $\overline{\Omega }=\Omega \cup \delta \Omega
$).{\ In the set }$\Omega \times I,$ $I${\ denoting an appropriate
finite time interval $I\subset
\mathbb{R}
,$ the relevant fluids }$\left\{ \rho =\rho
_{o}>0,\mathbf{V},p\right\} ,${\ i.e., respectively the (constant)
mass density, fluid velocity and pressure describing the fluid,
are assumed to be strong solutions of the so-called incompressible
Navier-Stokes equations (INSE) \cite{Ellero2005}. } The starting
point of the new statistical description here introduced is the
assumption that the fluid equations are stochastic in some sense
to be suitably specified. This permits us to define a
\emph{stochastic
decomposition }for all the relevant quantities, i.e., the fluid fields ${Z(%
\mathbf{r,}t\mathbb{)}\equiv }\left\{ \mathbf{V},p\right\} $ as
well the
volume force density $\mathbf{f}(\mathbf{r,}t\mathbb{)}$. Namely we let, $Z(%
\mathbf{r,}t\mathbb{)}\mathbb{=}\left\langle Z(\mathbf{r,}t\mathbb{)}%
\right\rangle +\delta Z(\mathbf{r,}t\mathbb{)}$ and $\mathbf{f}(\mathbf{r,}t%
\mathbb{)}\mathbb{=}\left\langle \mathbf{f}(\mathbf{r,}t\mathbb{)}%
\right\rangle +\delta \mathbf{f}(\mathbf{r,}t\mathbb{)}$, where $%
\left\langle {}\right\rangle $ is a suitable stochastic-averaging
operator.
Here, by proper definition of the operator $\left\langle {}\right\rangle ,$ $%
\left\langle Z(\mathbf{r,}t\mathbb{)}\right\rangle ,$ $\left\langle \mathbf{f%
}(\mathbf{r,}t\mathbb{)}\right\rangle $ and $\delta
Z(\mathbf{r,}t\mathbb{)}$ and $\delta
\mathbf{f}(\mathbf{r,}t\mathbb{)}$ represent respectively the
\emph{averaged parts} and the \emph{stochastic fluctuations}. In
the sequel the precise definition of the operator $\left\langle
{}\right\rangle $ is not required, however, we shall assume that
it commutes with all the
differential operators appearing in the previous fluid equations (i.e., $%
\frac{\partial }{\partial t},$ $\nabla $ and $\nabla ^{2}$). As a
consequence a suitable set of \emph{stochastic equations} can be
obtained.
In particular, the equations for the stochastic-averaged fields $%
\left\langle Z(\mathbf{r,}t\mathbb{)}\right\rangle $ read
\begin{equation}
\left\{
\begin{array}{c}
\nabla \cdot \left\langle \mathbf{V}\right\rangle =0, \\
\left\langle N\right\rangle \left\langle \mathbf{V}\right\rangle
+\left\langle \delta N\delta \mathbf{V}\right\rangle =0%
\end{array}%
\right.  \label{8}
\end{equation}%
(\emph{stochastic-averaged INSE}). Here the notation is standard \cite%
{Ellero2005}. Thus, $N$ is the nonlinear Navier-Stokes differential operator{%
\ }$N\mathbf{V}=\frac{\partial }{\partial
t}\mathbf{V}+\mathbf{V}\cdot \nabla \mathbf{V}+\frac{1}{\rho
_{o}}\left[ \nabla p-\mathbf{f}\right] -\nu \nabla
^{2}\mathbf{V},$ where $\mathbf{f}$ is {the force density,
}assumed
to be smooth real vector field. and $\rho _{o}$ and the kinematic viscocity $%
\nu $ are real positive constants, both to be considered
non-stochastic. Moreover, $\left\langle N\right\rangle $ and
$\delta N$ are respectively the
operators $\left\langle N\right\rangle \left\langle \mathbf{V}\right\rangle =%
\frac{\partial }{\partial t}+\left\langle \mathbf{V}\right\rangle
\cdot \nabla \left\langle \mathbf{V}\right\rangle +\frac{1}{\rho
_{o}}\left[
\nabla \left\langle p\right\rangle -\left\langle \mathbf{f}\right\rangle %
\right] -\nu \nabla ^{2}\left\langle \mathbf{V}\right\rangle $
and\textbf{\
\ }$\delta N\delta \mathbf{V}=\delta \mathbf{V}\cdot \nabla \delta \mathbf{V}%
+\frac{1}{\rho _{o}}\left[ \nabla \delta p-\delta
\mathbf{f}\right] -\nu
\nabla ^{2}\delta \mathbf{V,}$ while $\underline{\underline{\Pi _{R}}}(%
\mathbf{r,}t\mathbb{)}\equiv \left\langle \delta \mathbf{V}\delta \mathbf{V}%
\right\rangle $ is the Reynolds stress tensor.

\section{Deterministic IKT formulation for the stochastic-averaged INSE}

The discovery of the IKT for INSE \cite{Ellero2000} suggests us to
seek an inverse kinetic equation for the set of
stochastic-averaged equations defined by Eqs.(\ref{8}). In
particular we look for a \emph{Markovian inverse kinetic equation}
(IKE) of the Vlasov-type \cite{Tessarotto2008} which in Eulerian
form reads
\begin{eqnarray}
&&\left. L(\left\langle Z\right\rangle
)f(\mathbf{x},t;\left\langle
Z\right\rangle )=0\right. ,  \label{Liouville} \\
L(\left\langle Z\right\rangle )f &\equiv &\frac{\partial }{\partial t}f+%
\frac{\partial }{\partial \mathbf{x}}\cdot \left\{
\mathbf{X}(\left\langle Z\right\rangle )f\right\} ,
\end{eqnarray}%
(\emph{Eulerian IKE}). Here $f(\mathbf{x},t;\left\langle
Z\right\rangle )$ denotes the \emph{Eulerian local pdf} for
Eqs.(\ref{8}), which advances in
time the stochastic-averaged fluid fields $\left\langle Z(\mathbf{r,}%
t)\right\rangle $ and $L(\left\langle Z\right\rangle )$ is the
corresponding streaming operator$.$ We intend to construct
$f(\mathbf{x},t;\left\langle Z\right\rangle )$ and $L(\left\langle
Z\right\rangle )$ by imposing a suitable set of prescriptions.
Besides the requirement of validity of the fluid equations
(\ref{8}), these include in particular:

\begin{itemize}
\item $\mathbf{x=(r,v)}$ is the state vector spanning the restricted phase
space $\Gamma =\Omega \times V$ \cite{Tessarotto2008} and the vector field $%
\mathbf{X}$ has the form $\mathbf{\mathbf{X}}(\mathbf{\left\langle
Z\right\rangle )=}\left\{ \mathbf{v},\mathbf{F}(\left\langle
Z\right\rangle
)\right\} $, where $\mathbf{F}(\left\langle Z\right\rangle )\equiv \mathbf{F}%
(\mathbf{x},t;f,\left\langle Z\right\rangle )$ is an appropriate
mean-field force, to be assumed generally functionally dependent
on the local pdf.

\item By appropriate choice of the mean field force $\mathbf{F}$ and of the
local pdf, the moment equations can be prescribed in such a way to
satisfy identically Eqs.(\ref{8}). For this purpose we assume that
$f$ is a strictly positive, suitably smooth in $\Gamma \times I$
and summable both in the phase-space $\Gamma $ and in the velocity
space $V.$ In particular, we require that the (Shannon) entropy
integral
\begin{equation}
S(f)=-\int_{\Gamma }d\mathbf{x}f(\mathbf{x},t;\left\langle
Z\right\rangle )\ln f(\mathbf{x},t;\left\langle Z\right\rangle )
\label{entropy integral}
\end{equation}%
exists and that there results identically in $\Gamma \times I:$
\begin{equation}
\left\langle Z(\mathbf{r,}t)\right\rangle =\int d^{3}vG(\mathbf{x,}t)f(%
\mathbf{x},t;\left\langle Z\right\rangle ),  \label{moments}
\end{equation}%
where $\left\langle Z(\mathbf{r,}t)\right\rangle =1,\left\langle \mathbf{V}(%
\mathbf{r,}t)\right\rangle $ and $p_{1}(\mathbf{r,}t)$ are the
velocity
moments $G(\mathbf{x,}t)=1,\mathbf{v,}\frac{1}{3}\left\langle \mathbf{u}%
\right\rangle ^{2}.$ Moreover,
$\mathbf{u=v}-\mathbf{V}(\mathbf{r},t)$ is
the relative velocity and $p_{1}(\mathbf{r,}t)=P_{0}(t)+\left\langle p(%
\mathbf{r,}t)\right\rangle $ is\ the kinetic pressure and
$P_{0}(t)$ a smooth real function (pseudo-pressure) to be suitably
defined (see below).

\item The form of the pdf $f$ is chosen in such a way to satisfy the
principle of entropy maximization (PEM) \cite{Jaynes1957}, i.e.,
imposing the variational equation $\delta S(f)=0,$ with $\delta
^{2}S(f)<0,$ while
requiring that $f$ belongs to a suitable functional class $\left\{ f(\mathbf{%
x},t;\left\langle Z\right\rangle )\right\} ,$ to be determined
based solely on the information available on
$f(\mathbf{x},t;\left\langle Z\right\rangle ) $. \ In a turbulent
fluid this is manifestly provided by the knowledge of
the stochastic-averages of the fluid fields, $\left\langle Z(\mathbf{r,}%
t)\right\rangle $. In this case, imposing the constraints placed
by the moments (\ref{moments}) it is immediate to prove that PEM
yields necessarily\ as a particular equilibrium solution of the
inverse kinetic equation the local Maxwellian distribution
function (\emph{kinetic equilibrium)}
\begin{equation}
f_{M}(\mathbf{x},t;\left\langle Z\right\rangle )=\frac{1}{\left(
\pi \right) ^{\frac{3}{2}}v_{th}^{3}}\exp \left\{ -X^{2}\right\} ,
\label{Maxwellian}
\end{equation}%
where $X^{2}=\frac{\left\langle \mathbf{u}\right\rangle ^{2}}{vth^{2}},$ $%
v_{th}^{2}=2p_{1}/\rho _{o}$ and $\left\langle
\mathbf{u}\right\rangle \mathbf{=v}-\left\langle
\mathbf{V}(\mathbf{r},t)\right\rangle $. However, in principle
arbitrary non-Maxwellian initial kinetic distributions are
possible. These are potentially relevant, in particular, for
direct numerical simulations, in which the kinetic distribution
function is simulated numerically by means of test particles. In
such a case, in fact, small numerical errors may imply that
locally the kinetic distribution function may actually be
non-Maxwellian.

\item Eq.(\ref{Liouville}) implies the construction of a suitable classical
dynamical system, defined by a phase-space map%
\begin{equation}
\mathbf{x}_{o}\rightarrow \mathbf{x}(t)=T_{t,t_{o}}\mathbf{x}_{o},
\label{classical dynamical system}
\end{equation}%
where $T_{t,t_{o}}$ is the evolution operator
\cite{Tessarotto2006} generated by the initial-value problem
\begin{equation}
\left\{
\begin{array}{c}
\frac{d}{dt}\mathbf{x}=\mathbf{X}(\mathbf{x},t), \\
\mathbf{x}(t_{o})=\mathbf{x}_{o},%
\end{array}%
\right.   \label{in-2}
\end{equation}%
to be viewed as the \emph{Lagrangian} (or Langevin)
\emph{equations} for \ Eq. (\ref{Liouville}).

\item The equivalence between Eulerian and Lagrangian representations. In
fact the Eulerian IKE can also be represented in the equivalent \emph{%
Lagrangian form} \cite{Tessarotto2008}
\begin{equation}
J(\mathbf{x}(t),t)f(\mathbf{x}(t),t;\left\langle Z\right\rangle )=f(\mathbf{x%
}_{o},t_{o};\left\langle Z_{o}\right\rangle )\equiv f_{o}(\mathbf{x}%
_{o};\left\langle Z_{o}\right\rangle ),  \label{Eq.2}
\end{equation}%
(\emph{Lagrangian IKE}) where $f(\mathbf{x}(t),t;\left\langle
Z\right\rangle )$ is the Lagrangian representation of the pdf,
$\mathbf{x}(t)$ is the
solution of the initial-value problem (\ref{in-2}), $f_{o}(\mathbf{x}%
_{o};\left\langle Z_{o}\right\rangle )$ is{\ a suitably smooth
initial pdf and }$J(\mathbf{x}(t),t)=\left\vert \frac{\partial
\mathbf{x}(t)}{\partial
\mathbf{x}_{o}}\right\vert $ is the the Jacobian of the map $\mathbf{x}%
_{o}\rightarrow \mathbf{x}(t).$
\end{itemize}

The following theorem can be proven:\newline

\textbf{Theorem - Markovian IKT for the stochastic-averaged INSE
}\newline \emph{Let us assume that:} \emph{A}$_{1})$\emph{\
Eqs.(\ref{8}) admit a
smooth strong solution in }$\overline{\Gamma }\times I;$ \emph{A}$_{2}$\emph{%
) the} \emph{mean-field force}
$\mathbf{F}(\mathbf{x,}t;f;\left\langle Z\right\rangle )$
\emph{associated to the fluid fields} $\left\langle
Z\right\rangle $ \emph{reads:}%
\begin{equation}
\mathbf{F}(\mathbf{x},t;f,\left\langle Z\right\rangle )=\mathbf{F}_{0}+%
\mathbf{F}_{1},  \label{F}
\end{equation}%
\emph{where }$\mathbf{F}_{0}\mathbf{,F}_{1}$\emph{\ read respectively}%
\begin{equation}
\mathbf{F}_{0}\mathbf{(x,}t;f,\left\langle Z\right\rangle
)=\frac{1}{\rho
_{o}}\left[ \mathbf{\nabla \cdot }\underline{\underline{{\Pi }}}-\mathbf{%
\nabla }p_{1}+\left\langle \mathbf{f}_{R}\right\rangle \right] +\frac{1}{2}%
\left\langle \mathbf{u}\cdot \nabla \mathbf{V}\right\rangle +\frac{1}{2}%
\left\langle \nabla \mathbf{V\cdot u}\right\rangle \mathbf{+}\nu
\nabla ^{2}\left\langle \mathbf{V}\right\rangle \mathbf{,}
\label{F0}
\end{equation}%
\begin{equation}
\mathbf{F}_{1}\mathbf{(x,}t;f;\left\langle Z\right\rangle )=\frac{1}{2}%
\left\langle \mathbf{u}\right\rangle \left( \frac{1}{p_{1}}A\mathbf{+}\frac{1%
}{p_{1}}\mathbf{\nabla \cdot Q}-\frac{1}{p_{1}^{2}}\left[
\mathbf{\nabla
\cdot }\underline{\underline{\Pi }}\right] \mathbf{\cdot Q}\right) +\frac{%
v_{th}^{2}}{2p_{1}}\mathbf{\nabla \cdot }\underline{\underline{\Pi
}}\left( \frac{u^{2}}{v_{th}^{2}}-\frac{3}{2}\right) ,  \label{F1}
\end{equation}%
\emph{where }$A\equiv \frac{\partial }{\partial t}\left(
P_{0}(t)+\left\langle p\right\rangle \right) +\left\langle \mathbf{V}%
\right\rangle \mathbf{\cdot \nabla }\left( P_{0}(t)+\left\langle
p\right\rangle \right) ;$ \emph{A3) }$f(\mathbf{x,}t;\left\langle
Z\right\rangle )$ \emph{satisfies suitable initial and boundary
condition consistent with the initial-boundary value problem}
\emph{\ (\ref{8}) (see
Ref. \cite{Ellero2005}); A4) the initial pdf, }$f(\mathbf{x,}%
t_{o};\left\langle Z_{o}\right\rangle ),$ \emph{is suitably smooth
an strictly positive}; \emph{A5) }$f(\mathbf{x,}t;\left\langle
Z\right\rangle )$ \emph{admits the moments (\ref{entropy
integral}) and (\ref{moments}) as
well as }$\mathbf{Q}=\int d^{3}v\mathbf{u}\frac{u^{2}}{3}f$ \emph{and} $%
\underline{\underline{{\Pi }}}=\int d^{3}v\mathbf{uu}f,$ \emph{to
be assumed suitably smooth; A6) the pseudo-pressure }$P_{o}(t)$
\emph{is determined in
such a way that there results identically in }$I$%
\begin{equation}
\frac{\partial }{\partial t}S(f(t))=0  \label{H-theorem}
\end{equation}%
\emph{(constant H-theorem);} \emph{A7) in the fluid equations
(\ref{8}) and
in Eqs. (\ref{F0}) and (\ref{F1}) the Reynolds stress tensor} $\underline{%
\underline{\Pi _{R}}}\equiv \left\langle \delta \mathbf{V}\delta \mathbf{V}%
\right\rangle $ \emph{is considered a prescribed function of }$(\mathbf{r,}%
t) $\emph{.}

\emph{It follows that:\ }

\emph{B}$_{1}$\emph{)} \emph{the velocity-moment equations of IKE (\ref%
{Liouville}) evaluated for the weight functions} $G(\mathbf{x,}t)=\mathbf{v,}%
\frac{1}{3}u^{2}$ \emph{coincide with the stochastic INSE equations); B}$%
_{2} $\emph{) the Maxwellian pdf (\ref{Maxwellian}) is a
particular solution
of IKE. In particular, the local Maxwellian distribution function (\ref%
{Maxwellian}) is a particular solution of the IKE
(\ref{Liouville}) if an only if the fluid fields }$\left\{
\left\langle \mathbf{V}\right\rangle ,\left\langle p\right\rangle
\right\} $\emph{\ satisfy the
stochastic-averged Eqs.(\ref{8}). In such a case there results identically }$%
\mathbf{Q=0,}$\textbf{\ }$\underline{\underline{{\Pi }}}=\underline{%
\underline{0}}$\emph{; B}$_{3}$\emph{)
}$f(\mathbf{x,}t;\left\langle Z\right\rangle )$ \emph{is strictly
positive in }$\Gamma \times I.$\emph{\ Hence it is a probability
density.}\newline

PROOF - The proof is immediate. In fact: B$_{1}$) First, invoking Eqs.(\ref%
{F}),(\ref{F0}) and (\ref{F1}), it follows that the
velocity-moment
equations of IKE Eq.(\ref{Liouville}) for the weight functions $G(\mathbf{%
r,v,}t)=1,\mathbf{v,}\frac{1}{3}\mathbf{u}^{2}$ read respectively:
\begin{equation}
\nabla \cdot \left\langle \mathbf{V}\right\rangle =0,
\label{continuity equation}
\end{equation}%
\begin{equation}
\frac{\partial }{\partial t}\left\langle \mathbf{V}\right\rangle
+\left\langle \mathbf{V}\nabla \cdot \mathbf{V}\right\rangle \mathbf{+}\frac{%
1}{\rho _{o}}\left[ \mathbf{\nabla }\left\langle
p_{1}\right\rangle -\left\langle \mathbf{f}\right\rangle \right]
-\nu \nabla ^{2}\left\langle \mathbf{V}\right\rangle =0,
\label{average NS-2}
\end{equation}%
\begin{equation}
\nabla \cdot \left[ \left\langle \mathbf{V}\right\rangle
\left\langle p_{1}\right\rangle \right] +\left\langle
\mathbf{V}\right\rangle \cdot \nabla \left\langle
p_{1}\right\rangle =0  \label{average pressure eq.}
\end{equation}%
[which manifestly coincide with the fluid equations
Eqs.(\ref{8})]. Hence, also thanks to A6 the pdf advances in time
uniquely the stochastic-averaged fluid fields $\left\langle
Z(\mathbf{r,}t\mathbb{)}\right\rangle .$ B$_{2}$)
Second,.invoking A1 it is immediate to prove that $f_{M}(\mathbf{x}%
,t;\left\langle Z\right\rangle )$ is a particular solution of the
inverse
kinetic equation (\ref{Liouville})$.$ In fact, substituting (\ref{Maxwellian}%
) in the inverse kinetic equation (\ref{Liouville}) it follows:
\begin{equation*}
L(\left\langle Z\right\rangle )f_{M}=\left\{ \frac{\partial }{\partial t}%
\left\langle \mathbf{V}\right\rangle \mathbf{+v\cdot \nabla
}\left\langle
\mathbf{V}\right\rangle \right\} \mathbf{\cdot }\frac{\left\langle \mathbf{u}%
\right\rangle \rho _{o}}{\left\langle p_{1}\right\rangle
}f_{M}\mathbf{+}
\end{equation*}%
\begin{equation}
\left. +\left\{ \frac{\partial }{\partial t}\ln \left\langle
p_{1}\right\rangle \mathbf{+v\cdot \nabla }\ln \left\langle
p_{1}\right\rangle \right\} \left\{ \frac{\left\langle u\right\rangle ^{2}}{%
\left\langle v_{th}^{2}\right\rangle }-\frac{3}{2}\right\}
f_{M}-\right.
\end{equation}%
\begin{equation*}
-\mathbf{F}(\mathbf{x},t;f_{M},\left\langle Z\right\rangle )\cdot \frac{%
\left\langle \mathbf{u}\right\rangle \rho _{o}}{\left\langle
p_{1}\right\rangle }f_{M}+f_{M}\frac{\partial }{\partial
\mathbf{v}}\cdot \mathbf{F}(\mathbf{x},t;f_{M},\left\langle
Z\right\rangle )=0.
\end{equation*}%
Thanks to Eqs.(\ref{F}),(\ref{F0}) and (\ref{F1}), there results:%
\begin{equation*}
L(\left\langle Z\right\rangle )f_{M}=\left\{ \frac{\partial }{\partial t}%
\left\langle \mathbf{V}\right\rangle \mathbf{+v\cdot \nabla
}\left\langle \mathbf{V}\right\rangle +\frac{1}{\rho _{o}}\left[
\nabla \left\langle p_{1}\right\rangle +\left\langle
\mathbf{f}\right\rangle \right] -\right.
\end{equation*}%
\begin{equation*}
\left. -\left\langle \mathbf{u}\cdot \nabla \mathbf{V}\right\rangle \mathbf{-%
}\nu \nabla ^{2}\left\langle \mathbf{V}\right\rangle \right\} \cdot \frac{%
\left\langle \mathbf{u}\right\rangle \rho _{o}}{\left\langle
p_{1}\right\rangle }f_{M}
+f_{M}\nabla \cdot \left\langle \mathbf{V}%
\right\rangle =0
\end{equation*}%
which implies Eqs.(\ref{8}). \ Instead, if we assume that in
$\Gamma \times I,$ $f\equiv f_{M}(\mathbf{x},t;\left\langle
Z\right\rangle )$ is a particular solution of the inverse kinetic
equation, which fulfills identically the constraint equation
(\ref{H-theorem}), it follows that the fluid fields $\left\langle
\mathbf{V}\right\rangle ,\left\langle p\right\rangle $ are
necessarily solutions of the INSE equations. If, instead, $f\equiv
f_{M}(\mathbf{x},t;\left\langle Z\right\rangle )$ is a particular
solution of Eq.(\ref{Liouville}), thanks to B$_{1}$ it follows
that the stochastic INSE equations are necessarily fulfilled.
B$_{3}$)
Finally, due to Eq.(\ref{Liouville}) the entropy production\ rate reads%
\begin{eqnarray}
\frac{\partial }{\partial t}S(f(t)) &=&-\int_{\Gamma }d\mathbf{x}\frac{%
\partial }{\partial t}f(\mathbf{x,}t;\left\langle Z\right\rangle )\left[
1+\ln f(\mathbf{x,}t;\left\langle Z\right\rangle )\right] = \\
&=&-\int_{\Gamma }d\mathbf{x}f(\mathbf{x,}t;\left\langle Z\right\rangle )%
\frac{\partial }{\partial \mathbf{v}}\cdot \mathbf{F}(\mathbf{x}%
,t;f,\left\langle Z\right\rangle ).  \notag
\end{eqnarray}%
Hence, thanks to Eqs. (\ref{F}),(\ref{F0}),(\ref{F1}), the constraint Eq.(%
\ref{H-theorem}) requires%
\begin{equation}
\frac{\partial }{\partial t}S(f(t))=-\frac{3}{2}\int_{\Omega }d^{3}\mathbf{r}%
\frac{1}{P_{0}(t)+\left\langle p\right\rangle }\left[
A\mathbf{+\nabla \cdot
Q}-\frac{1}{p_{1}}\left[ \mathbf{\nabla \cdot }\underline{\underline{\Pi }}%
\right] \mathbf{\cdot Q}\right] =0,
\end{equation}%
which delivers an ordinary differential equation for the
pseudo-pressure. Assuming that the fluid fields $\left\langle
\mathbf{V}\right\rangle
,\left\langle p\right\rangle $ and the moments $\mathbf{Q}$ and $\underline{%
\underline{{\Pi }}}$ are suitably smooth, this equation can always
be fulfilled in a finite time interval $I.$ As a direct
consequence, it follows that $f(\mathbf{x,}t;\left\langle
Z\right\rangle )$ is manifestly strictly positive in $\Gamma
\times I$.

\section{Conclusions}

In this paper a statistical model of hydrodynamic turbulence has
been formulated by means of a deterministic inverse kinetic theory
(IKT) for the stochastic-averaged INSE. Basic aspects of the new
theory are: A) that the
IKT satisfies exactly the relevant fluid equations, represented by Eqs.(\ref%
{8}), while B) the pdf advances in time uniquely the
stochastic-averaged fluid fields $\left\langle Z\right\rangle $.
As a main consequence, the theory fulfills identically a closure
condition. In fact, there exists, by construction, a subset of the
moment equations which is closed and manifestly coincides with the
same set of stochastic fluid equations. The result B) holds
provided the Reynolds stress tensor can be considered prescribed
(to be identified, for example, with\emph{\ }a suitably weighted
velocity-space integral of the local pdf). While the form of the
tensor remains in principle unspecified, this limitation shall be
lifted in an accompanying paper \cite{Tessarotto2008-aa}. The
present theory displays several remarkable new features. In
particular, unlike customary statistical approaches, it is based
on the introduction of the local position-velocity joint
probability density function (local pdf). In addition, the kinetic
equation advancing in time the pdf is a Markovian Vlasov-type
kinetic equation which admits a straightforward equivalent
Lagrangian formulation. Under suitable prescriptions the form of
this equation can be proven to be unique \cite{Tessarotto2006}.
Further key property is that the kinetic equation admits in
general, besides a local Maxwellian kinetic equilibrium, also
non-Gaussian solutions. In our view the theory provides a
convenient setting both for the investigation of theoretical
aspects of turbulence theory. These include - besides the
mathematical formulation of turbulence problems - several
important physical applications, such as, for example,
the connection with Fokker-Planck statistical descriptions \cite%
{Tessarotto2008-aa} and the Lagrangian dynamics of particles in
turbulent flows and its generalization to incompressible thermofluids \cite%
{Tessarotto2008}.

\section*{Acknowledgments}
Work developed (M.T.) in the framework of the MIUR (Italian
Ministry of University and Research) PRIN Research Program
``Modelli della teoria cinetica matematica nello studio dei
sistemi complessi nelle scienze applicate'' and the European COST
action P17 (M.T). The partial of the GNFM (National Group of
Mathematical Physics) of INDAM (National Institute of Advanced
Mathematics, Italy) (M.T. and P.N.) and of the Deutsche
Forschungsgemeinschaft via the project EL503/1-1 (M.E.) is
acknowledged.

\section*{Notice}
$^{\S }$ contributed paper at RGD26 (Kyoto, Japan, July 2008).




\newpage
\begin{thebibliography}{BIBTEX}
\bibitem{Tessarotto2008-aa} M. Tessarotto, M.Ellero, D.Sarmah and P.Nicolini,
\textit{Fokker-Planck Kinetic description of small-scale fluid
turbulence for classical incompressible fluids}, contributed
paper, 26th RGD, Kyoto, Japan, July 2008.

\bibitem{Kolm} A.\ N.\ Kolmogorov, C.\ R.\ Acad.\ Sci.\ URSS, \textbf{30},
301 (1941).

\bibitem{experiment} H.\ L.\ Grant, R.\ W.\ Stewart, and A.\ Molliet, J.\
fluid Mech., \textbf{12}, 241, (1962); U.~Frisch and P.\ Sulem,
Phys.\ Fluids, \textbf{27}, 1921, (1984);

\bibitem{Monin1975} A.S. Monin and A.M. Yaglom, \textit{Statistical Fluid
Mechanics}, Vol. 1 and 2, MIT Press (1975); U. Frisch,
\textit{Turbulence}, Cambridge University Press (1995).

\bibitem{Pope2000} S.B. Pope, \textit{Turbulent flows}, Cambridge University
Press, p.463 (2000).

\bibitem{Chen1989} H.Chen, S. Chen and R.H. Kraichnan, Phys. Rev.Lett
\textbf{63}, 2657 (1989); S.P. Pope, Prog. Energy Cobist. Sc. \textbf{11}%
,119 (1985); A.M. Polyakov, Phys. Rev. E \textbf{52}, 6183 (1995).

\bibitem{Aringazin} A.K. Aringazin und M.I. Mazhitov, Int. J. Mod. Phys.
\textbf{B 18}, 3095 (2004).

\bibitem{Friedrich} R. Friedrich, Phys. Rev. Lett. \textbf{90}, 084501,
(2003); A. Baule and R. Friedrich, Phys. Rev. E \textbf{71},
026101 (2005).

\bibitem{Sreenivasan1997} K.R. Sreenivasan and R.A. Antonia, Ann.Rev.Fluid
Mech \textbf{29},435 (1997).

\bibitem{Ellero2000} M. Ellero and M. Tessarotto, Bull. Am Phys. Soc.
\textbf{45 }(9), 40 (2000).

\bibitem{Ellero2005} M. Ellero and M. Tessarotto, Physica A \textbf{355},
233 (2005); M. Tessarotto and M. Ellero, Proc. 24th RGD, Bari,
Italy (July 2004), Ed. M. Capitelli, AIP Conf. Proc. \textbf{762},
108 (2005).

\bibitem{Tessarotto2006} M. Tessarotto and M. Ellero, Physica A \textbf{373}%
, 142 (2007); arXiv: physics/0602140; M. Tessarotto and M. Ellero,
Proc. 25th RGD (International Symposium on Rarefied gas Dynamics,
St. Petersburg, Russia, July 21-28, 2006), Ed. M.S. Ivanov and
A.K. Rebrov (Novosibirsk Publ. House of the Siberian Branch of the
Russian Academy of Sciences), p.1001; arXiv:physics/0611113
(2007).

\bibitem{Tessarotto2008} Marco Tessarotto, Claudio Cremaschini and Massimo
Tessarotto, \textit{Lagrangian dynamics of incompressible
thermofluids},
contributed paper, 26th RGD, Kyoto, Japan, July 2008; M. Tessarotto, \textit{%
Inverse kinetic theory for classical and quantum fluids}, ibid.

\bibitem{Jaynes1957} E.T. Jaynes, Phys. Rev. \textbf{106}, 620 (1957).

\end{thebibliography}
\end{document}